**Sub-1-Angstrom-Resolution Imaging Reveals Phase Contrast Transition in Ice I$_h$ Caused by Basal Stacking Faults**

Jingshan S. Du[1*], Suvo Banik[2], Lehan Yao[1], Shuai Zhang[1,3], Subramanian K. R. S. Sankaranarayanan[2,4], James J. De Yoreo[1,3*]

[1]Physical Sciences Division, Pacific Northwest National Laboratory, Richland, WA 99352, United States

[2]Center for Nanoscale Materials, Argonne National Laboratory, Lemont, IL 60439, United States

[3]Department of Materials Science and Engineering, University of Washington, Seattle, WA 98195, United States

[4]Department of Mechanical and Industrial Engineering, University of Illinois, Chicago, IL 60607, United States

[*]Correspondence should be addressed to james.deyoreo@pnnl.gov (J.D.Y.) or jingshan.du@outlook.com (J.S.D.)

**Abstract:** Phase-contrast transmission electron microscopy (TEM) of hexagonal ice (I$_h$) along [0001] sometimes shows a honeycomb-like pattern, often interpreted as individual oxygen columns in single crystals. Here, we show that this pattern commonly arises from intrinsic basal stacking faults instead. A translational boundary separating domains of comparable thickness, with an in-plane offset of (⅔$a_1$ + ⅓$a_2$), produces this honeycomb-like contrast. Stacking domains translated in nonequivalent directions yields patterns resembling cubic ice (I$_c$) along [111] but with a 3-fold symmetry. We imaged this structure at a record-breaking line resolution of 89 picometers, finer than the O-H covalent bond length. These findings highlight the defect tolerance of ice's molecular packing and clarify the structural relationships among hexagonal, stacking-disordered, and cubic ice phases. This resolution milestone opens new avenues for characterizing subtle structural perturbations of water in the solid state.

**Main Text**

Ice is a crucial family of crystals and a key model system for understanding crystallization mechanisms and intermolecular interactions [1,2]. The relative weakness of hydrogen bonds between water molecules renders the tetrahedral structural motifs prone to distortion [1]. This flexibility leads to diverse crystal symmetries under varying pressure and temperature conditions [3] and facilitates the formation of crystal imperfections even near ambient conditions [4]. However, studying the molecular arrangement in ice at the atomic or molecular scale, particularly how it is perturbed at defects and interfaces, has been very challenging [4]. This difficulty arises mainly because ice is highly susceptible to damage from high-energy imaging sources, water has a low scattering cross-section in many irradiations, and preparing ice samples compatible with high-resolution imaging instruments is nontrivial [5].

Recently, breakthroughs in cryogenic high-resolution transmission electron microscopy (HRTEM) have enabled direct phase-contrast imaging of type-I ice, yielding atomic-scale information. Kobayashi *et al.* reported the first images of this kind from condensed ice crystallites on TEM grids mounted on a liquid-nitrogen cryogenic sample holder [6]. Low-dose cryogenic HRTEM has been further used to study kinetically trapped ice deposition from the low-pressure residue vapor inside the TEM column directly onto lacey carbon or graphene [7,8] or ice thin films derived from the vitrification procedure used in biological sample preparation [9,10]. Our team developed a new method, termed cryogenic liquid-cell TEM (CRYOLIC-TEM), that has enabled atomic-resolution imaging of various defects and interfacial structures in ice frozen from liquid water under near-thermodynamic conditions [5]. This approach protects high-quality, crystalline ice thin films from the high-vacuum environment and electron damage by encasing the sample between amorphous carbon membranes. The ice crystals can withstand HRTEM imaging conditions for minutes, and dynamical processes such as bubble nucleation, movement, and shrinkage within the crystals were observed *in situ* at lattice resolution [5].

In HRTEM images of hexagonal ice (type $I_h$; this paper uses the $I_h$(c) notation [11,12], space group $P6_3/mmc$, no. 194) along the *c* axis, [0001], two types of phase-contrast patterns are frequently observed. The dominant contrast is a hexagonal dot array (Fig. 1a), and a less commonly seen pattern shows a honeycomb structure (Fig. 1b). In the latter pattern, the arrangement of the bright dots appears to coincide with the oxygen atomic columns (Fig. 1c), thus has conventionally been used as a strong evidence to the arrangement of water molecules in ice $I_h$ single crystals [6,13]. The flipping between the two patterns is commonly attributed to contrast transfer, which varies with sample thickness and defocus [6,13]. In particular, it is believed that the transition between the two patterns in the same HRTEM image directly reflects a variation in crystal thickness [13].

Here, we show that the transition observed in the experiment is often impermissible by contrast transfer in a single crystal. Atomic-resolution HRTEM images of ice thin films were obtained using the CRYOLIC-TEM approach, as reported earlier [5]. An aberration-corrected FEI Titan HRTEM (corrected for the image-forming lenses) and a Gatan Metro 300 direct electron detector operating in electron-counting mode allowed for the acquisition of high-quality phase-contrast data. The methods are detailed in Appendix A. When a transition from the "dot array" (Pattern 1) to "honeycomb" (Pattern 2) occurs in adjacent areas in the same lattice, the bright dot positions in Pattern 1 remain bright in Pattern 2 (Fig. 1d, along orange dashes), contrary to the phase contrast reversal caused by a variation of sample thickness or defocus (*vide infra*). In fact, the intensity of the same row of bright-dot features is on the same level in both regions (Fig. 1e). A similar transition can also occur within the same area over time. Time-series, spatially registered images recorded under a constant electron flux are used to demonstrate the transition from Pattern 2 to 1 (Fig. 1f). By tracking the evolution of individual features, we found that the transition occurs through the emergence or disappearance of half of the bright dots in the honeycomb pattern (purple disks), while the other half stay bright (blue disks).

The observed transition behavior is inconsistent with image contrast transfer theories based on thickness–defocus variation. Dynamical TEM image simulations show that although a transition between a "dot array" and a "honeycomb" pattern can occur by changing the sample thickness and defocus for a single crystal, the bright dots in the former always turn dark in the latter, and the "honeycomb" dots emerge at the interstitial positions (Fig. 2a). Additionally, the "honeycomb" pattern is usually much weaker than the dominant "dot array" pattern, with only about half the maximum intensity (see also Fig. S1 in Supplemental Material [14]). Moreover, reversing the contrast in neighboring areas within the same thin-film sample is very difficult. In fact, it is almost impossible to flip the contrast by solely changing the thickness of the thin-film sample below 100 nm (Fig. 2a, vertical direction). Transition through defocus adjustment also requires at least 20–30 nm of change (Fig. 2a, horizontal direction). In experiments, this typically relies on the unevenness of the supporting membrane, which may shift the sample's vertical position. Yet, atomic force microscopy (AFM) shows that the height variation of grid-supported carbon membranes over a 100 × 100 nm area is only on the order of 1 nm (Fig. 2b), which is an order of magnitude smaller. Tilting the sample to yield a height difference matching the horizontal length scale, as shown in Fig. 1d, would require an unrealistically large tilt (e.g., 45°), which was not present in the experiments. Possible defocus variation due to field curvature is also at the picometer level in HRTEM, making it highly unlikely to meet the requirements (see Appendix B). Lastly, contrast changes of this kind should not break the 6-fold symmetry. This cannot account for the symmetry breaking observed in the experiments, in which half of the spots exhibit different intensities and anisotropies than the other half (Fig. 1b and f). This symmetry breaking is also evident in phase-contrast images reported in the literature [6,13]. Even when accounting for sample or beam tilting, the 3-fold symmetry seen in the experiments is unlikely to form.

Another possibility for introducing such symmetry breaking is to have only a few monolayers of molecules, thereby breaking the unit cell. This phenomenon has been observed during the monolayer-to-bilayer transition in 2D materials such as graphene [15]. In the ice lattice, a similar behavior can be observed for ½-, ¾-, and 1-unit-cell configurations along the $c$ axis. Multislice HRTEM simulations at near-zero defocus show that a transition from a simple hexagonal dot array to a symmetry-broken honeycomb pattern, and then to a full honeycomb, occurs as one monolayer of water molecules is added at a time (Fig. 2c and d). Here, the images are essentially a direct projection of the atomic potential field from the oxygen atoms; hence, this behavior is not permissible with thicker films (contrary to a recent report [13]). The ice thin films prepared by CRYOLIC-TEM and other methods are typically on the 10-to-100-nm scale and are much thicker than 2–4 monolayers [5]. As such, we conclude that the symmetry breaking of projected atomic potentials in 2D ice layers is not responsible for the contrast transition.

For crystals thicker than a few monolayers, the observed symmetry breaking must involve vertically stacked domains separated by translational domain boundaries. The basal molecular packing in type-I ice resembles the stacking correlation in hexagonal close-packed (hcp) and face-centered cubic (fcc) metals. Indeed, the ABAB… stacking (each letter represents a water bilayer) in hexagonal ice along the *c* axis can form intrinsic stacking faults and introduce C-layers that are observed in the ABCABC… stacking of cubic ice along [111] (Fig. 3a). Such an intrinsic stacking fault separates two domains that have an in-plane translational vector ($\frac{2}{3}a_1 + \frac{1}{3}a_2$) [16]. There are two sets of equivalent translations that can occur, depending on the direction: one results in a BC-domain on AB, and the other yields an AC-domain on AB (Fig. 3b). Note that for each set, two types of intrinsic stacking faults ($I_1$ and $I_2$) are permitted, although $I_2$ is more commonly observed in ice [8-10].

We simulated HRTEM images of an ice crystal with a total thickness of 25 unit cells, with varying ratios of AB- and BC-domains. The presence of either stacking fault results in a 3-fold-symmetric, honeycomb-like pattern, and the intensity of the two sets of dots in the honeycombs is tunable by a combination of the thickness ratio of the two domains and the defocus value (Fig. 3c). This pattern is consistent with the transitional behavior observed in the experiments. The transition from simple "dot arrays" to the symmetry-broken "honeycombs" only requires the formation of an intrinsic basal stacking fault, which typically requires partial dislocations or local cubic-like embryos during nucleation, and is known to occur easily in type-I ice [17-19]. Indeed, past results have shown that the formation energy of such stacking faults is < 1 meV/atom [19-21], which is significantly lower than the scale of thermal fluctuation at 0 °C ($k_BT \approx 23.5$ meV) and even at −180 °C ($k_BT \approx 8$ meV), i.e., the temperature at which the samples are frozen during imaging. As such, perturbations during or after crystallization, such as shear or electron-beam effects, can readily induce the formation of such structures.

When AB, BC, and AC stacking modes are present simultaneously, a more complex pattern forms. This arrangement is resolved by phase-contrast CRYOLIC-TEM imaging of an ice thin film, achieving a record-high line resolution of 89 picometers (Fig. 3d and e). By spatially averaging the repeating patterns across the image, a 3-fold-symmetric pattern was identified: three groups of bright dots with different intensities overlay within the same repeating unit cells (Fig. 3f). This pattern matches the structural model of vertically stacked AB, BC, and AC domains along the $c$ axis (Fig. 3g). Notably, all three sets of patterns coexist in local areas, although their relative intensities and distortions vary slightly at different locations (Fig. 3h). This observation further confirms that they originate from separate crystal layers stacked in the vertical direction.

The aforementioned pattern can be reliably replicated *in silico* by modeling a 3-layer crystal of translational domains. In this case, multislice HRTEM simulation of BC|AB|AC domains with thicknesses of 11, 8, and 6 unit cells (totaling 25 unit cells) displays the same 3-fold-symmetry pattern observed in the experiment, and the three sets of bright dots have distinct intensity levels (Fig. 3i). From a structural standpoint, such a crystal is sometimes called "stacking disordered" type-I ice ($I_{sd}$) in the literature [18,22]. Most experimentally produced cubic type-I ice ($I_c$) is widely understood to be structurally imperfect, exhibiting a mixture of $I_h$ and $I_c$ domains along with stacking faults, and may fall into the category of $I_{sd}$ [23,24]. Only recently has phase-pure cubic ice been successfully isolated through conversion from hydrogen hydrates [25,26]. A pure $I_c$ crystal forms when A, B, and C molecular layers occur consecutively and evenly, creating ABCABC… stacking along its [111] direction. In this case, all bright dots in the pattern become equivalent, restoring a 6-fold symmetry in the phase contrast (Fig. 3j).

To further understand the conversion between single-crystalline and stacking-fault-bearing ice $I_h$ crystals, coarse-grained molecular dynamics (MD) simulations were performed using a machine-learned bond order potential (ML-BOP) model of water [27]. This model correctly captures the thermodynamic properties of water and the ice-water phase boundary, in good agreement with experiments [27,28]. We began by modeling an ice $I_h$ single crystal (i.e., AB-type; with periodic boundary conditions) and relaxing it *in silico* at 240 K to reach equilibrium (Fig. 4a). Here, each water molecule was modeled as a bead. We then applied a shear strain to the crystal by displacing a section of water molecules (see Appendix A.5 for details). The displacement increases linearly along the $c$ axis, producing the largest shifts for molecules at the center of the box. This setup causes local shear deformations in the crystal, which are relevant to the mechanical or electron-beam conditions ice encounters during freezing or imaging. This section was then relaxed for $2 \times 10^5$ timesteps (0.2 ns) before the entire crystal was allowed to relax. The initial partial freezing helps to stabilize the structure and slow down the transformation. Upon lattice disruption, portions of the crystal amorphized and then recrystallized into a defect-bearing structure (Figs. 4b to d).

The transformation involves two stages. First, up to ~0.08 ns, the crystalline section in the middle (pink shades, Fig. 4b) gradually moved in-plane from AB-type to BC-type positions. This is evidenced by a reduction in the mean $y$ positions of the beads in this layer, while those in $x$ and $z$ were relatively stable (Fig. 4e). During this stage, the crystallization of amorphous beads mainly extends existing hexagonal crystals, while stacking faults with localized cubic-like packing remain a minority (Figs. 4f). The transformation rate also decreases, as evidenced by a slowdown in both the reduction of amorphous beads (Fig. 4f) and the total system energy change (Fig. 4g).

As this middle layer settles into the BC-type positions, a second stage of crystallization begins (from ~0.08 to 0.2 ns). Here, we observed rapid formation of cubic-like layers from the amorphous regions (Figs. 4c, d, and f). These local structures are intrinsic basal stacking faults that connect AB- and BC-type domains. This structural evolution results in a transition from the "dot array" phase-contrast pattern in pure hexagonal ice (Fig. 4a) to a complex pattern with three independent sets of dots (Fig. 4d), as significant amounts of AB-, BC-, and AC-type stacking are present in the resulting structure (62.5%, 12.5%, and 25% of bilayer stackings), an observation consistent with the experimental results. The entire crystal is finally allowed to relax above 0.2 ns. Comparing the faulted crystal to a control group where no shear was introduced, the total system energy converges to a very similar value after ~0.5 ns and can be considered to be at equilibrium. The equilibrium energies of the sheared (faulted) crystal and the phase-pure hexagonal ice crystal (control) are indistinguishable from each other within thermal fluctuations (Fig. 4h), consistent with previous discussions.

This simulation provides strong evidence that basal stacking fault formation can be spontaneous once the energy barrier is overcome, and that the faulted structures have formation energies that are negligible relative to thermal fluctuations. In scenarios such as water freezing or electron irradiation, local shear stresses—driven by mass transport, membrane interaction, or kinetic energy transfer—can readily initiate this process, facilitating the dynamic formation and migration of stacking faults by breaking and reforming local molecular bonds. Crucially, our simulations reveal that the crystalline layers bounding the amorphous regions undergo in-plane translation prior to recrystallization. This temporal sequence suggests that the transient amorphous phase functions as a lubricated slip zone, thereby uncoupling lattice planes and facilitating plastic flow [29]. On the mesoscale, this sequence—localized amorphization, layer translation, and faulted recrystallization—can be characterized as dislocation migration mediated by a transient phase transformation. Here, the amorphous region acts as an extended, mobile dislocation core that lowers the friction stress for glide, a mechanism analogous to the shear-induced amorphization and triboepitaxy observed in covalent materials [30,31].

In summary, we have shown that the "honeycomb" pattern observed along [0001] of hexagonal ice can originate from translational domains separated by intrinsic basal stacking faults; therefore, they do not necessarily reflect the oxygen atomic columns in single crystals. These patterns have several distinct characteristics:

1. The "honeycomb" and "dot array" patterns can transition from one to the other both spatially and temporally.
2. In a transition, the bright dots in the "dot array" pattern remain bright, which directly contradicts the contrast reversal caused by thickness–defocus variation in single crystals.
3. The intensity of half of the dots in the "honeycomb" pattern can vary, resulting in a pattern with only 3-fold symmetry.

This pattern originates from vertically stacked domains with an in-plane offset of ($\frac{2}{3}a_1 + \frac{1}{3}a_2$) separated by intrinsic basal stacking faults. If multiple domains with different offset directions are present, a pattern similar to that of cubic ice along [111], but with only 3-fold symmetry, appears. MD simulations further show spontaneous fault development by breaking and reforming local molecular stacking on the basal planes, resulting in heterogeneous structures with free energies indistinguishable from those of phase-pure hexagonal ice.

This work, based on the CRYOLIC-TEM approach we developed earlier [5], has now crossed the 1-angstrom resolution threshold. It indicates that, for the first time, the line resolution of electron micrographs of ice is finer than the O-H covalent bond length in water molecules (~0.96 Å in liquid and ~1.00 Å in ice $I_h$) [32,33], a pivotal landmark that brings imaging capabilities for ice to the (sub)molecular level. This unprecedented resolution allowed us to elucidate the complex phase-contrast patterns formed by translational domain boundaries in this work, and it may further unlock the potential to experimentally map local lattice distortions associated with proton ordering [34], to elucidate molecular rearrangements at defects and interfaces [4], and to directly observe host-guest interactions in clathrate hydrates [35,36].

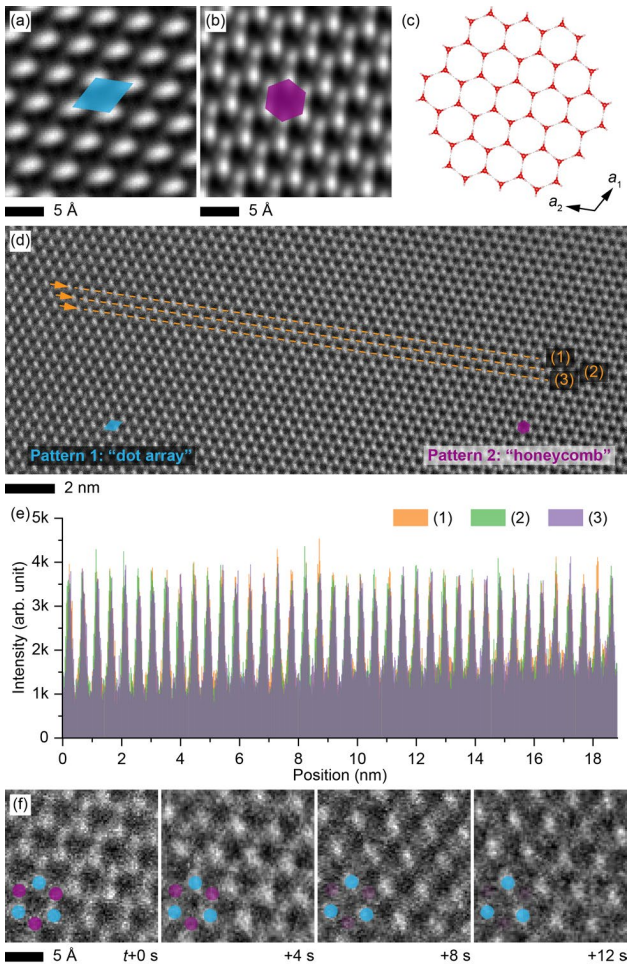

**Figure 1. Phase contrast transition in HRTEM of an ice $I_h$ thin film along [0001].** (a, b) Spatially averaged HRTEM showing "dot array" (a) and "honeycomb" patterns (b) from the same crystal. The elongation of the dot shapes is due to the slight crystal tilt [5]. Colored polygons serve as visual guides. (c) Structural illustration of ice $I_h$ viewed along [0001]. (d) Frequency domain gamma corrected (FDGC) HRTEM showing the spatial transition between two patterns. Orange dashes indicate rows of bright dots. (e) Intensity profiles along the orange dashes in (d). (f) Time-series spatially registered FDGC-HRTEM showing the temporal transition from Pattern 2 to Pattern 1 in a local region. Blue and purple disks indicate the persistent and diminishing dots in a honeycomb, respectively. Electron flux at detector: 60 e $Å^{-2}$ $s^{-1}$.

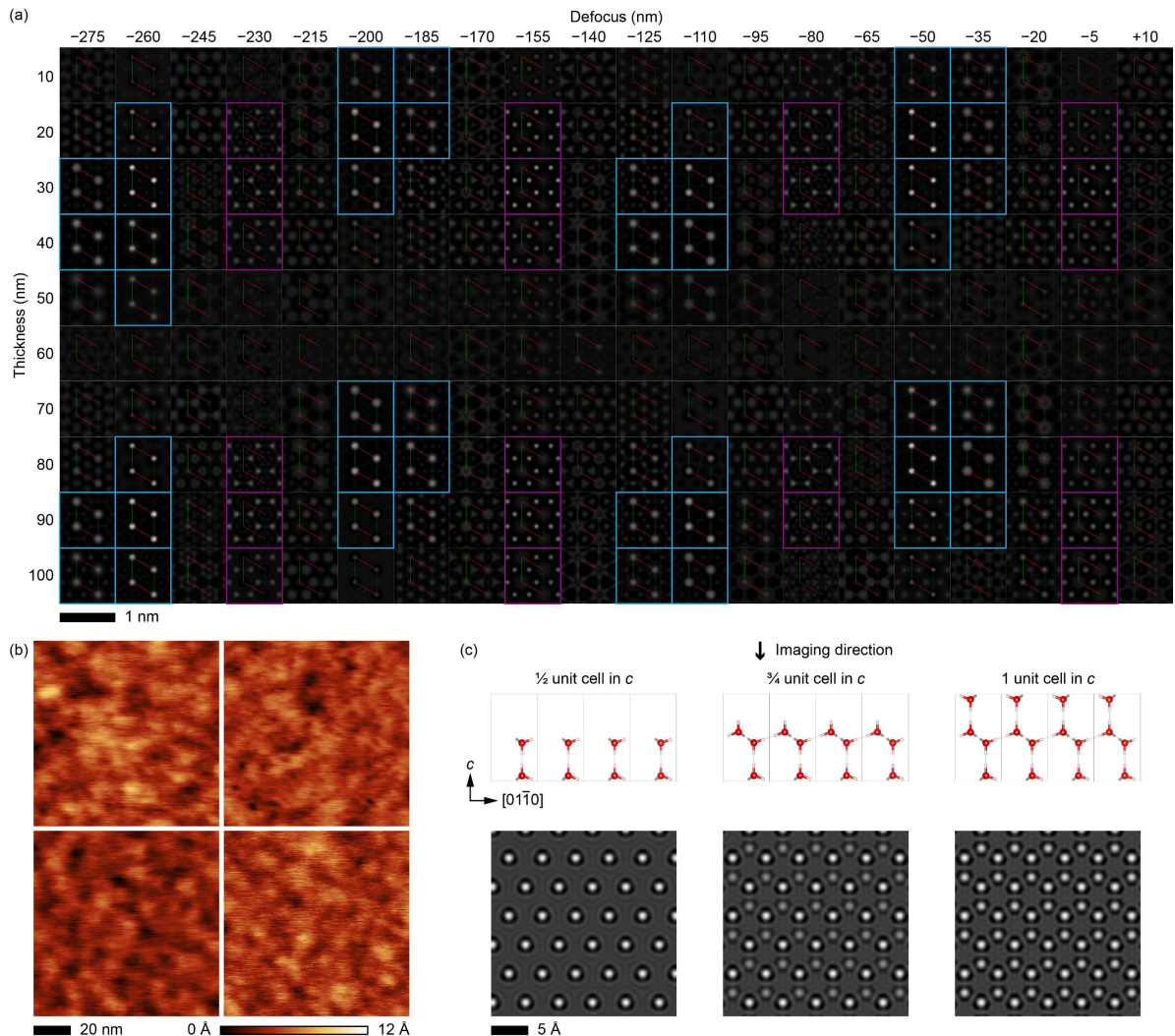

**Fig. 2. Effect of thickness–defocus variation on contrast transfer.** (a) Dynamical simulation of HRTEM images of ice $I_h$ along [0001] by varying the sample thickness (by row) and defocus (by column). Blue and purple boxes indicate strong patterns in the "dot array" and "honeycomb" configurations, respectively. Only images showing one distinct pattern and a maximum intensity of at least 25% of the global maximum of this matrix are identified. The unit cell boundary is labeled in each image. Image intensities are globally normalized. (b) AFM height maps of the grid-supported carbon membranes used in CRYOLIC-TEM at four different locations (sampling interval: 2 × 4 Å). (c) Structural models (upper row) and multislice HRTEM simulation along [0001] at a defocus of +4 nm (lower row) for 2D ice with only a few molecular layers in the $c$ direction.

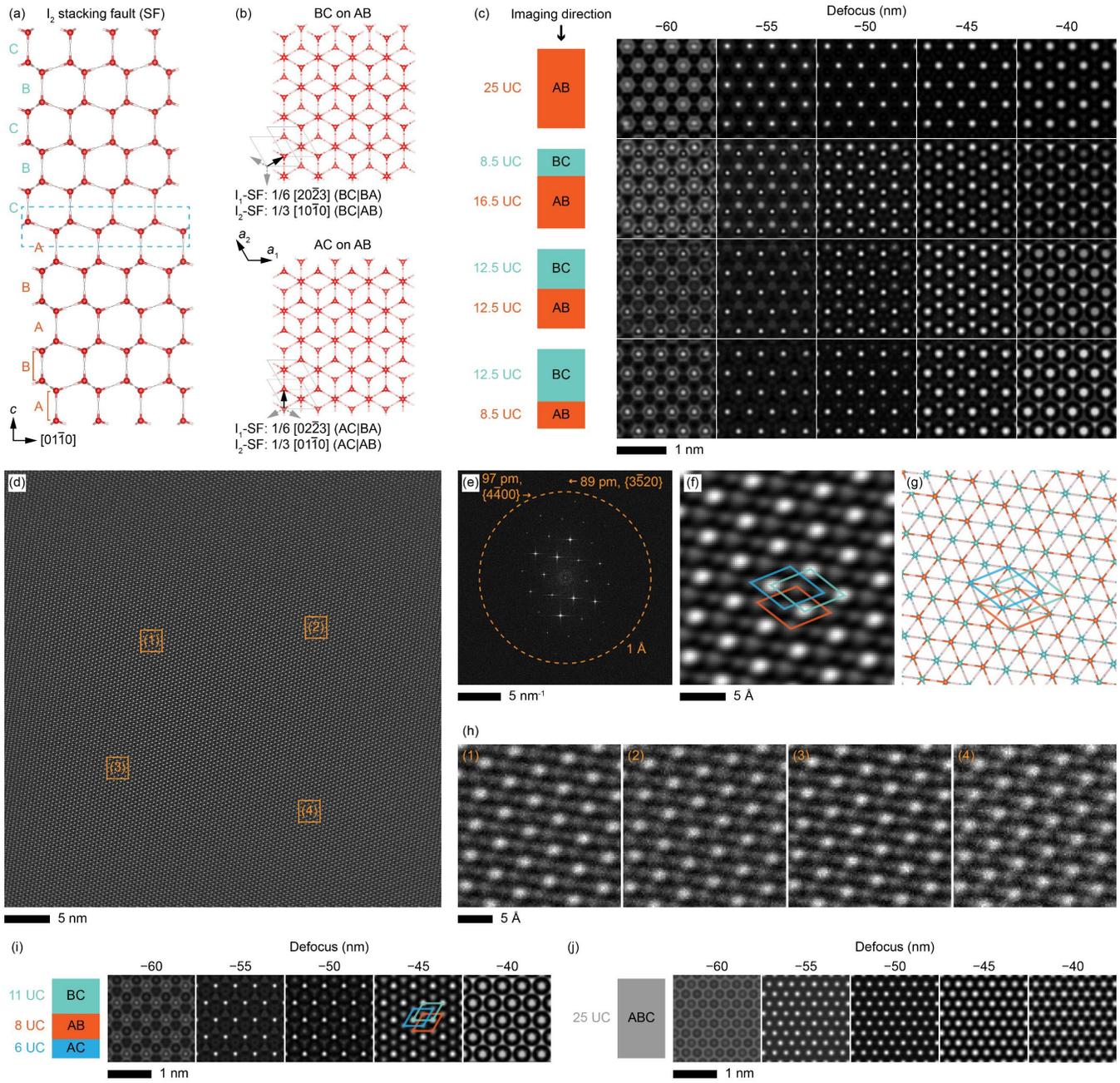

**Fig. 3. Stacked translational domains causing symmetry-broken phase contrast.** (a) Structural illustration of an I$_2$-type stacking fault in hexagonal ice. The dashed box indicates the fault plane. (b) Illustrations of two translational equivalents for the domain configurations; each allows both I$_1$- and I$_2$-type stacking faults. (c) Multislice TEM simulation of two stacked translational domains of different thicknesses (in unit cells, UC) and defocus values along [0001]. (d) FDGC-HRTEM of an ice crystal with all three types of domains (AB, BC, and AC) stacked in the vertical direction. (e) Fourier transform of (d). (f, g) Spatially averaged HRTEM (f) and corresponding structural model (g). (h) Zoomed-in FDGC-HRTEM from locations labeled in (d). (i, j) Multislice TEM simulation of three stacked translational domains along [0001] (i) and cubic ice (I$_c$) along [111] (j). In (f), (g), and (i), unit cells from different crystal layers are illustrated in three colors.

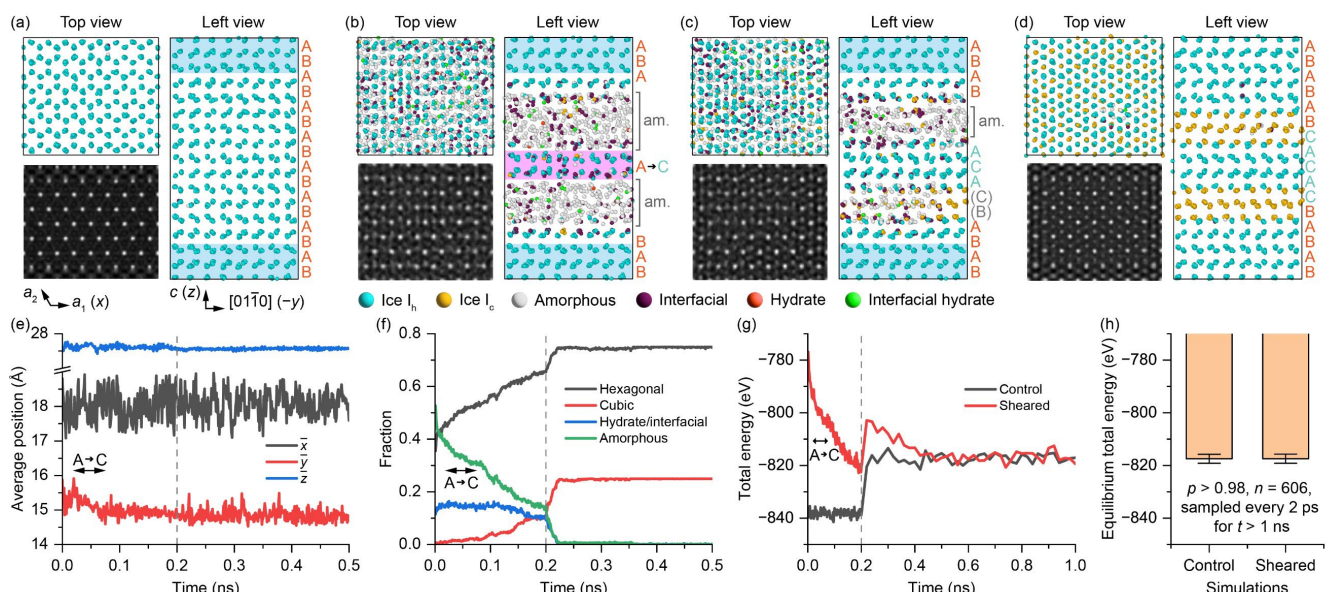

**Fig. 4. Molecular dynamics simulation of stacking faults generated by shear deformation.** (a) Structure and multislice TEM simulation (lower left) of a phase-pure ice $I_h$ crystal (periodic boundary conditions) post-relaxation. (b–d) Structure and multislice TEM simulation showing stacking fault development at 0.05 (b), 0.15 (c), and 0.25 ns (d) after a shear was introduced. Panels (a) to (d) share the same axis definitions and color schemes below them. The region in pink shades indicates the middle layer undergoing an in-plane translation. Regions in blue shades were frozen in the first 0.2 ns. (e) Average position of the middle layer [pink shades in (b)] versus time. (f) Temporal evolution of local structures. (g) Temporal evolution of total system energy. (h) Comparison of the equilibrium total system energies.

**Supplemental Material**
The Supplemental Material includes Supplemental Figures S1–S2 (PDF).

**Acknowledgments**
We thank D. Peng, L. Kovarik, and Y. Xu of Pacific Northwest National Laboratory (PNNL) and B. K. Miller (Gatan, Inc.) for helpful discussions. Microscopy and analysis were supported by the U.S. Department of Energy (DOE) Office of Science (SC) Basic Energy Sciences (BES) Division of Materials Science and Engineering, Synthesis and Processing Sciences program (FWP 67554) at PNNL (J.J.D.Y.) and a Seed Laboratory-Directed Research and Development (LDRD) project at PNNL Physical and Computational Sciences Directorate (J.S.D.). Molecular dynamics simulations were supported by the Data, Artificial Intelligence, and Machine Learning at Scientific User Facilities program under the Digital Twin Project at Argonne National Laboratory (S.K.R.S.S.). A portion of this research was performed on a project award (60789) from the Environmental Molecular Sciences Laboratory at PNNL (J.S.D. and J.J.D.Y.). Work at the Center for Nanoscale Materials was supported by the U.S. DOE SC BES under Contract No. DE-AC02-06CH11357. J.S.D. acknowledges a Washington Research Foundation Postdoctoral Fellowship. PNNL is a multiprogram national laboratory operated for the DOE by Battelle under Contract DE-AC05-76RL01830.

**Notes**
The authors declare no competing financial interest.

**Data Availability**
The data that support the findings of this article are openly available [37].

**Appendix A: Methods**

*1. Cryogenic liquid-cell transmission electron microscopy (CRYOLIC-TEM)*
Atomic-resolution CRYOLIC-TEM data of hexagonal ice along the [0001] zone axis were obtained following a previously published report [5]. Briefly, deionized water was sandwiched between two amorphous carbon membranes and loaded onto a Gatan 626 liquid nitrogen cryo-transfer holder. The liquid water film was frozen in the cryo transfer station using liquid nitrogen before the holder was inserted into the TEM column. An FEI Titan Environmental TEM [extreme-brightness Schottky field-emission gun (X-FEG), 300 kV] equipped with a CEOS double-hexapole aberration corrector (CETCOR) for the image-forming lenses and a Gatan Metro 300 direct electron detector was used for imaging.

*2. TEM image processing*
Spatial averaging of HRTEM was performed by (1) selecting a small patch in the source image (e.g., 100 by 100 pixels); (2) calculating the cross-correlation between this patch and the entire source image; (3) finding peak coordinates and intensities in the cross correlation; (4) averaging patches that are centered around peak coordinates with intensities above a certain threshold.

Frequency domain gamma correction (FDGC) was implemented by raising the magnitude to a power while keeping the phase unchanged in the frequency space. That is, for a Fourier transform
$$F(k,l) = |F(k,l)|e^{j\theta(k,l)}, \tag{A1}$$
the corrected version becomes
$$F^*(k,l) = |F(k,l)|^\gamma e^{j\theta(k,l)}. \tag{A2}$$
This method is a direct, nonlinear filtering operation conceptually related to homomorphic signal processing, which also uses nonlinearities in a transformed domain to alter signal characteristics. When $\gamma > 1$, the low-amplitude, broadband components in the frequency space, including noise, are effectively suppressed. A key advantage of this method is that it reweighs all frequency components without completely discarding any spectral information. This preserves the overall spectral shape and is less prone to artifacts that aggressive masking approaches can introduce. In this manuscript, $\gamma = 1.3$ was used in all cases.

*3. Structural modeling and TEM image simulation*
The crystal structure of ice $I_h$ was derived from the $I_h(c)$ modification in the literature (space group $P6_3/mmc$, no. 194; COD #9015208) [11,12]. Dynamical TEM simulation was performed using the ReciPro package [38] (https://seto77.github.io/ReciPro/). The simulation assumes a parallel beam condition, a beam energy of 300 keV ± 0.8 eV, $C_s = 4$ μm, and $C_c = 1.4$ mm. Multislice TEM simulation [39] was performed with the QSTEM package [40] (https://www.physik.hu-berlin.de/en/sem/software/software_qstem) using the same TEM parameters.

*4. Atomic force microscopy (AFM)*
AFM images of the carbon membrane supported on carbon meshes were acquired using an Asylum Research Cypher ES AFM in tapping mode with BudgetSensors Multi75AI-G AFM probes (Aluminum-coated, resonance frequency ~ 75 kHz, spring constant ~ 3 N/m) under ambient conditions. The images were corrected only for scanline artifacts using linear slopes prior to presentation. Image processing and quantification used the Gwyddion package [41] (http://gwyddion.net/).

## 5. *Molecular dynamics (MD) simulation*

The coarse-grained MD simulations were performed based on the ML-BOP model of water [27] in LAMMPS [42] (with periodic boundaries), starting from a hexagonal configuration. Specifically, two end-cap slabs of thickness $d_z = 10$ Å were defined at the bottom and top of the cell and treated as fixed boundaries during deformation, while the interior atoms were allowed to relax. The whole system was first relaxed under the *NPT* ensemble at $T = 240$ K and $P = 0$ for 0.10 ns. After relaxation, the end caps were frozen, and a symmetric same-direction shear ramp was applied by displacing interior atoms along $+y$ with a linear ramp in $z$. Specifically, letting the mobile span be $z \in [z_1, z_2]$ with midpoint $z_0 = (z_1 + z_2)/2$, the imposed maximum displacement is $\Delta = \gamma/2 \, (z_2 - z_1)$, and the displacement field is

$$u_y(z) = \begin{cases} \Delta \frac{z-z_1}{z_0-z_1}, & z \in [z_1, z_0], \\ \Delta \frac{z_2-z}{z_2-z_0}, & z \in [z_0, z_2], \end{cases} \text{with } u_x = u_z = 0, \quad (A3)$$

which yields equal-magnitude, oppositely signed shear gradients $\frac{\partial u_y}{\partial z} = \pm\gamma$ on the two halves while keeping the displacement direction $+y$ from both ends. This process introduces a localized shock into the system, mimicking the injection of energy under realistic probing conditions. The deformed structure was then relaxed via a short *NVT* "shock-settle" applied only to the interior atoms for 0.20 ns to remove high-frequency transients, followed by the release of the end caps and a long *NPT* relaxation at $T = 240$ K and $P = 0$ for 20.0 ns. Immediately before the final *NPT* stage, another control simulation was performed using the same protocol, except that no shear was applied to the system. Multislice TEM simulation based on MD results was performed with two supercells in the $z$ direction.

## Appendix B. Estimation of Potential Defocus Changes due to Off-Axial Aberrations

In electron microscopy, especially at a high magnification [e.g., under HRTEM conditions with a *ca.* 100 × 100 nm field of view (FOV) or even smaller], the defocus change across the FOV resulting from aberrations is negligible [43,44]. The scale of the curvature radius of the focal plane, known as Petzval field curvature in optics, of a thin magnetic lens (first-order approximation), can be estimated as [45,46]

$$R = \frac{f}{2}, \quad (B1)$$

where *f* is the focal length. For multiple lenses (1 to *i*), we can estimate the total curvature by

$$\frac{1}{R_i} = \sum_{j=1}^{i} \frac{1}{R_j}. \quad (B2)$$

In HRTEM, the focal lengths of the objective and projection lenses are typically on the order of 1–5 mm. Therefore, the radius of the curved field is on the scale of 1 mm. This yields a vertical offset on the scale of only 1 pm at a location 50 nm from the detector center. The off-axial astigmatism (also a third-order aberration) may further modify the wavefront shape in the meridional and sagittal directions on the same order of magnitude [43,44]. As such, the phase field in HRTEM is sufficiently flat, and the defocus variation caused by field curvature is safely negligible.

17

**Supplemental Material for**

**Sub-1-Angstrom-Resolution Imaging Reveals Phase Contrast Transition in Ice $I_h$ Caused by Basal Stacking Faults**

Jingshan S. Du[1*], Suvo Banik[2], Lehan Yao[1], Shuai Zhang[1,3], Subramanian K. R. S. Sankaranarayanan[2,4], James J. De Yoreo[1,3*]

[1]Physical Sciences Division, Pacific Northwest National Laboratory, Richland, WA 99352, United States
[2]Center for Nanoscale Materials, Argonne National Laboratory, Lemont, IL 60439, United States
[3]Department of Materials Science and Engineering, University of Washington, Seattle, WA 98195, United States
[4]Department of Mechanical and Industrial Engineering, University of Illinois, Chicago, IL 60607, United States
[*]Correspondence should be addressed to james.deyoreo@pnnl.gov (J.D.Y.) or jingshan.du@outlook.com (J.S.D.)

This PDF file contains:
    Fig. S1–S2

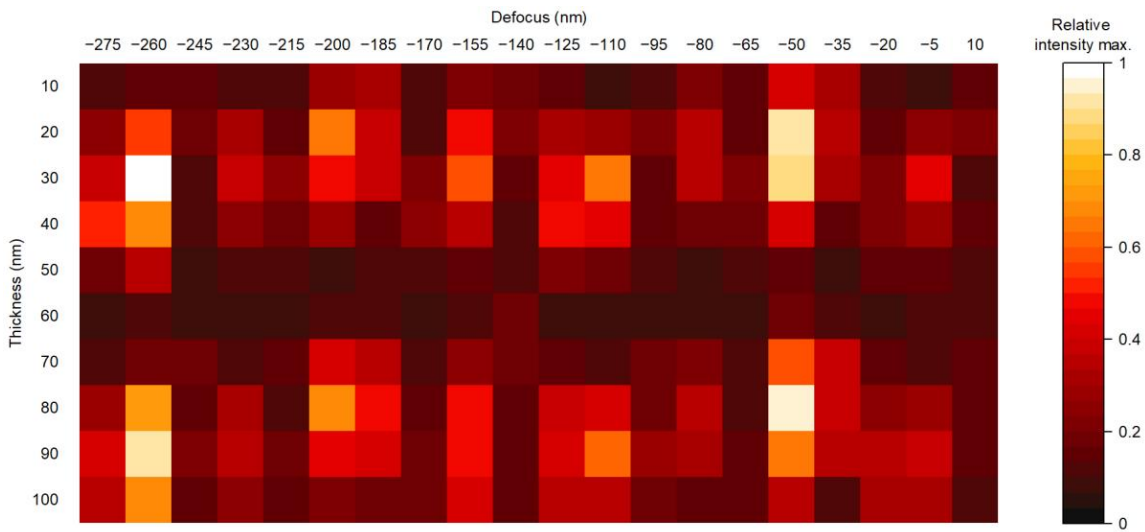

**Fig. S1. Heatmap showing the relative intensity maxima of dynamically simulated TEM images of ice I$_h$ along [0001] with different thickness and defocus values.** See Fig. 2a for the corresponding images.

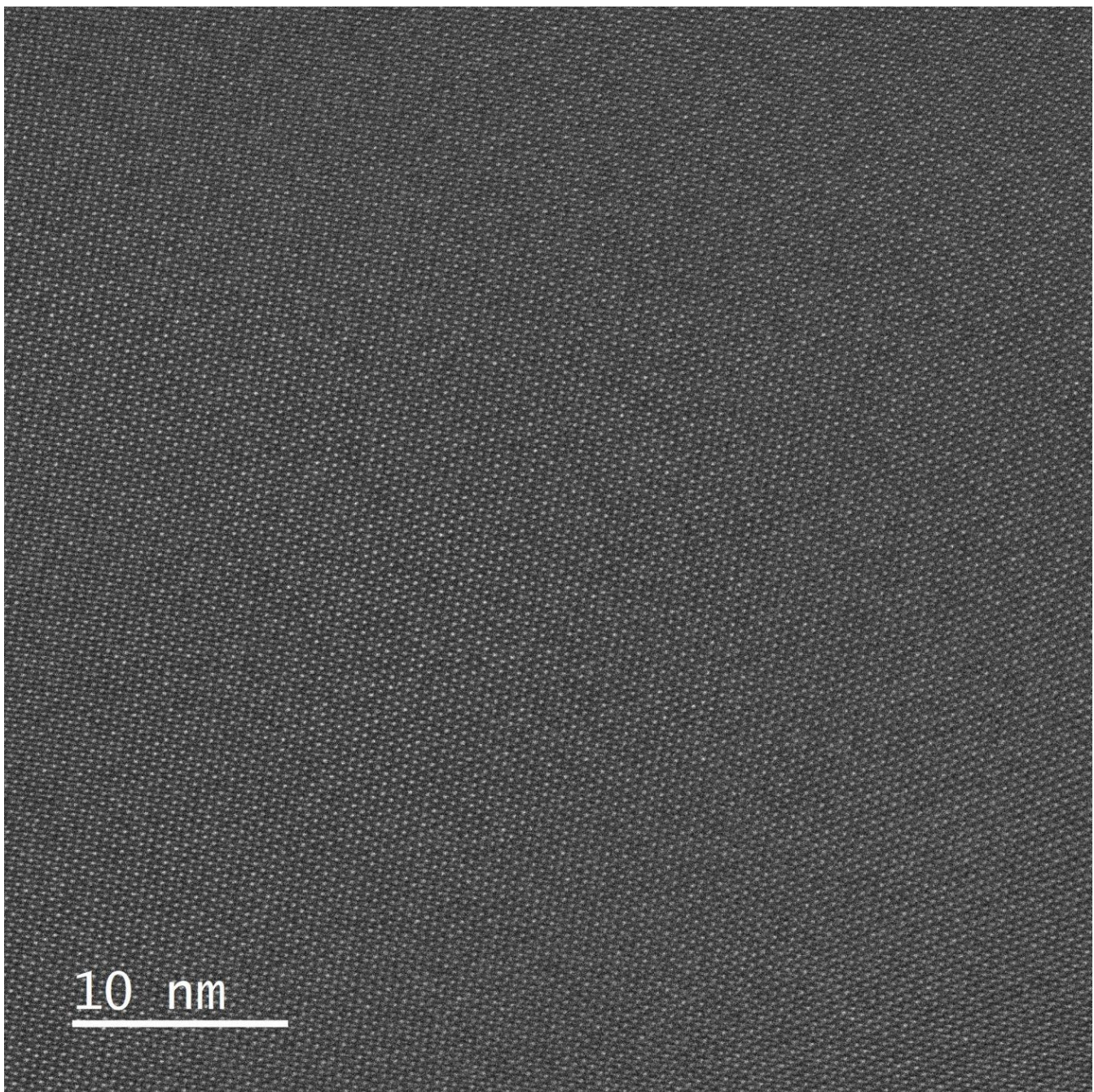

**Fig. S2. Unprocessed HRTEM source image for Fig. 3d.**